# On the LRD of the Aggregated Traffic Flows in High-Speed Computer Networks

G. Millán, *Member, IEEE*

*Abstract*— This paper studies and analyses the behavior of the Long-Range Dependence in network traffic after classifying traffic flows in aggregated time series. Following Differentiated Services architecture principles, the generic Quality of Service applications that requirements and use the transport control protocol, a basic classification criterion of time series is established. Using the fractal theory, the resulting time series are analyzed. The Hurst exponent is estimated and used as a measure of traffic burstiness and Long-Range Dependency in each traffic class. The traffic volume per class is also measured. The study uses traffic traces collected at the core switch at the Electric Engineering Department at Universidad de Santiago de Chile in different periods of network activity.

*Keywords*—Aggregated series, Differentiated Services (DiffServ) architecture, Hurst exponent (*H*), Long-Range Dependence (LRD), Quality of Service (QoS).

## I. Introducción

La diversidad de los requisitos de calidad de servicio (QoS) tanto de los servicios actuales como emergentes obliga a la red (Internet) a diferenciar los flujos de tráfico de forma tal que se ofrezca un nivel de QoS adecuado. Una de las soluciones más prometedoras propuestas por el Internet Engineering Task Force (IETF) es la arquitectura de servicios diferenciados (DiffServ) [1], la cual agrega tráfico en una cantidad limitada de clases de servicio según los objetivos de QoS. Este paradigma de tráfico plantea un desafío para el análisis y la caracterización del tráfico Internet; y aún cuando una multitud de estudios se centran en la caracterización general del tráfico en Internet, los efectos de la agregación de tráfico en las clases aún no son claros. ¿Será una clase de tráfico particular responsable del comportamiento que se observa en [2]?, ¿la agregación afecta las características del tráfico en ráfagas en nodos y enlaces de Internet?

En este trabajo se estudia la dependencia de largo alcance (LRD) por constituir la principal propiedad de la fractalidad de los flujos de tráfico, para dar respuesta a las dos preguntas anteriores.

Este estudio se lleva a cabo con trazas de tráfico capturadas con Wireshark [3] en el switch L3 que enlaza nodo central de la red corporativa de la Universidad de Santiago de Chile con el Departamento de Ingeniería Eléctrica. Todas las muestras se analizan aplicando un criterio de clasificación de tráfico basado en un enfoque de campo múltiple de la arquitectura DiffServ. De la misma forma, las características temporales de las trazas se estudian recurriendo al software MATLAB registrando los tiempos de arribo al sniffer y el tamaño de los paquetes como pares ordenados.

G. Millán, Facultad de Ingeniería y Tecnología, Universidad San Sebastián, Puerto Montt, Chile, ginno.millan@uss.cl.

## II. La Arquitectura de Servicios Diferenciados

La arquitectura DiffServ proporciona un método que intenta garantizar QoS en redes de gran tamaño como Internet. Los servicios diferenciados generan un comportamiento diferente por salto en cada router o switch, inspeccionando la cabecera de cada paquete IP para decidir cómo llevar a cabo el envío de ese paquete. Toda la información necesaria para esta decisión viaja en la cabecera, que por si misma no puede disponer su propio manejo, la cual presenta ciertos campos, que sirven para decidir envíos basados en las políticas de QoS adoptadas [4].

La redefinición del byte TOS de la cabecera IPv4 dio lugar al campo DiffServ usando los 6 bits más significativos [5], lo que aumenta la flexibilidad y las opciones. Los dos bits menos significativos llamados ECN (Explicit Congestion Notification) se usan para control de flujo. El modelo creado de esta forma se llama DSCP (Differentiated Services Code Point) y es compatible con IP Precedence (primeros 3 bits del campo TOS original de IPv4) para efectos de migraciones [5].

En la terminología DiffServ, el comportamiento de reenvío asignado a un DSCP se denomina PHB (Per-Hop Behavior). El PHB define la precedencia de reenvío de un paquete marcado en relación con otro tráfico del sistema con DiffServ. Dicha precedencia determina si el sistema IP QoS o DiffServ reenvía o descarta el paquete. Para un paquete reenviado, cada router o switch DiffServ que el paquete encuentra en la ruta hasta su destino aplica el mismo PHB. Una excepción ocurre toda vez que otro sistema DiffServ modifica el DSCP [5], [6].

Existen cuatro tipos de PHB con los valores del DSCP [5]

1. **Selector de Clase PHB**, con los 3 bits menos significativos a 0 del DSCP se obtiene compatibilidad con IP Precedence.
2. **PHB por defecto (BE)**, con los 3 bits más significativos del IP Precedence/DSCP se obtiene Best-effort.
3. **PHB Assured Forwarding (AF)**, con los 3 bits más significativos del DSCP puestos a 001, 010, 011 o 100 se usa para garantizar ancho de banda.
4. **PHB Expedite Forwarding (EF)**, con los 3 bits más significativos del DSCP puestos a 101 se usa para brindar un servicio de bajo retardo.

EF se usa para crear servicios que requieren baja pérdida, retardo y fluctuación de fase reducidos, y un ancho de banda garantizado. AF, que consta de cuatro clases, se usa para crear servicios con ancho de banda mínimo garantizado y diferentes niveles de tolerancia al retardo y la pérdida.

En IPv6 se utiliza un campo específico, Traffic Class, de 8 bits, con los seis primeros para DiffServ, y los dos últimos para ECN.

## III. CARACTERIZACIÓN DE TRÁFICO DE RED

El conocimiento de las características del tráfico de red en su conjunto, y en particular de los tráficos agregados, es relevante para permitir una asignación y gestión adecuada de los recursos de red para ayudar a la ingeniería de tráfico, el control de tráfico y congestión, y para especificar servicios de forma realista. Para este estudio el análisis se basa en la teoría de series temporales de carácter fractal, ya que estudios recientes relacionados con la caracterización y el modelado del tráfico de red apuntan a la presencia de autosimilitud y LRD. Esta última propiedad afecta directamente a los elementos resaltados anteriormente con un fuerte impacto en las colas y la naturaleza de la congestión [7].

### A. Tráfico Fractal

La autosimilitud expresa la invariancia de una estructura de datos independientemente de la escala en que se analizan. Desde una perspectiva de tráfico de red, la autosimilitud expresa una noción de estallido, es decir, no existe una longitud natural para una ráfaga y la estructura en ráfagas de tráfico se mantiene para diferentes escalas temporales. Ejemplo de proceso que exhibe autosimilitud y LRD es $X(t)$, un proceso estocástico autosimilar asintóticamente de segundo orden, con parámetro Hurst dado en $0.5 < H < 1$, es decir [8]

$$\lim_{m \to \infty} \gamma(k) = 0.5\sigma^2[(k+1)^{2H} - 2k^{2H} + (k-1)^{2H}]. \quad (1)$$

$X(t)$ posee las siguientes dos propiedades

1. **Dependencia de largo alcance**. La función de autocorrelación $r(k)$ decae hiperbólicamente; $r(k)$ es no sumable.
2. **Decaimiento lento de la varianza**. La varianza de una serie agregada a nivel $m$, $X_k^{(m)}$, de $X_k$ obtenida de $X(t)$, es decir

$$X_k^{(m)} = \frac{1}{m} \sum_{i=km-m+1}^{km} X_i, \quad \text{con } k, m \in \mathbb{N}, \quad (2)$$

cumple $\text{var}(X_k^{(m)}) \sim \text{var}(X_k) m^{-\beta}$, $\beta = 2 - 2H$, $0 < \beta < 1$.

El exponente de Hurst se emplea para medir la LRD y es un indicador de la impulsividad del tráfico (la impulsividad aumenta con H). Si ½ < $H$ < 1 entonces se observa persistencia infinita, y si 0 < $H$ < ½ el comportamiento no persistente ocurre mientras $H$ = ½. Existen varios métodos para estimar el parámetro $H$ [2]. En tanto que los métodos para probar la varianza, el estadístico R/S o el periodograma son de naturaleza gráfica, el estimador de Whittle proporciona un método analítico para estimar $H$. La prueba de varianza de este estudio se basa en la propiedad de la varianza que se descompone lentamente y $H$ se obtiene mediante $H = 1 - \beta/2$ con un diagrama log-log de (var ($X^{(m)}$), var ($X$) $m^{-\beta}$).

### B. Recolección de Muestras y Preparación

Las muestras se obtienen de un router principal ubicado en la troncal de la red corporativa de la Universidad de Santiago de Chile usando la herramienta software Wireshark [3]. Wireshark considera que se trata de un flujo unidireccional de paquetes de una fuente a un destino y registra en cada información de tiempo de entrada, campos tales como las direcciones IP de origen y destino, los números de puerto, el identificador de protocolo, las interfaces de entrada y salida y el número de paquetes y bytes enviados. La recolección de tráfico se lleva a cabo durante diferentes períodos de tiempo y varios días. Estos periodos de tiempo se eligieron reflejando los niveles típicos de actividad del Departamento de Ingeniería Eléctrica (baja: de 08 p.m. a 08 a.m., media: de 08 a.m. a 10 a.m. y de 03 p.m. a 05 p.m.; alta: de 10 a.m. a 01 p.m. y de 05 a 08 p.m.) Cada hora de tráfico la muestra se filtra a través de la interfaz de salida de acuerdo con los criterios que se detallan, para los diferentes intervalos temporales (100 ms, 500 ms, 1 s y 10 s).

### C. Criterio de Clasificación de Tráfico

Debido a razones económicas y técnicas, la definición de un criterio de clasificación de tráfico es una tarea subjetiva. Por ejemplo, para tipos de tráfico idénticos, un cliente puede estar dispuesto a pagar más que otros para obtener una mejor calidad de servicio. Además, cuando un criterio se basa en encabezados de paquetes TCP / UDP / IP, la fragmentación de paquetes, el cifrado de paquetes y el uso de puertos de aplicación negociados o no registrados dificultan la clasificación. Luego, un criterio de clasificación debe ser lo suficientemente genérico y robusto para ser fácilmente adoptado e implementado.

La mayoría de los criterios sugieren distintas clases de tráfico UDP y TCP para que las aplicaciones no reactivas y reactivas no compitan por los mismos recursos. Algunos sugieren además que la duración de los flujos, la velocidad de transmisión y las características de tamaño de paquete también deben considerarse [9]. También se propone un método de clasificación basado en los requisitos de la aplicación QoS, como la sensibilidad a la demora, la pérdida o los retardos [10].

Teniendo en cuenta los aspectos anteriores y el tipo de servicio (TOS) propuesto para aplicaciones clásicas [10], el criterio de clasificación está orientado a la agregación de tráfico que puede asignarse fácilmente a una arquitectura de QoS basada en clases. De esta manera como primer enfoque, el proceso de clasificación distingue el tráfico TCP del UDP y, a continuación, se tienen en cuenta los requisitos genéricos de las aplicaciones. Un proceso de filtrado basado en reglas más detalladas para diferenciar aplicaciones específicas o patentadas (por ejemplo como NetMeeting y Cisco IP/TV, y muchas otras aplicaciones de unidifusión o multidifusión) queda abierto. Así, las clases de tráfico resultantes son las siguientes

1. **Clase 1:** tráfico TCP resultante de aplicaciones interactivas como Telnet o SSH.
2. **Clase 2:** tráfico TCP sensible a pérdidas y rendimiento que resulta de aplicar protocolos de transferencia masiva.
3. **Clase 3:** tráfico HTTP.
4. **Clase 4:** tráfico proveniente de protocolos de enrutamiento o administración (TCP/UDP).
5. **Clase 5:** tráfico UDP genérico.
6. **Clase 6:** tráfico de aplicaciones con puertos transitorios o puertos UDP no cubiertos por las **clases 4 y 5**.

No hay una asignación directa entre las clases definidas y los PHB DiffServ dado que tal mapeo dependería de las políticas de servicio administrativas y contractuales. No obstante una posible coincidencia podría ser: **Clases 1** y **3** respaldadas por AF PHB de alta prioridad; **Clase 4** por EF PHB; **Clases 2** y **5** por BE o AF PHB de baja prioridad y **Clase 6** por EF o AF dependiendo de la relevancia a la diversidad de aplicaciones.

## IV. Análisis Estadístico de Datos

### A. Volúmenes de Tráfico

La Tabla I presenta el porcentaje de tráfico que contribuye, por clase, a la carga total en el router. Los resultados muestran que la única clase (**Clase 6**) cuyo contenido no está claramente identificado representa una pequeña cantidad de tráfico, lo cual demuestra la amplitud del criterio de clasificación propuesto.

Como se espera las **Clases 2** y **3** que principalmente incluyen tráfico granulado y web, respectivamente, contribuyen en gran medida a la carga global del router. Los resultados mostrados también dan cuenta de la existencia de una correlación entre los porcentajes de paquetes y bytes para todas las clases, aunque las **Clases 2** y **5** indican presencia de tamaños de paquetes grandes y pequeños, respectivamente.

TABLA I
VOLUMEN DE TRÁFICO PARA CADA UNA DE LAS SEIS CLASES DE TRÁFICO DEFINIDAS PARA EL ANÁLISIS

| Clase | Bytes (%) | Paquetes (%) |
|---|---|---|
| 1 | 2.31 | 2.37 |
| 2 | 26.40 | 18.95 |
| 3 | 67.04 | 68.41 |
| 4 | 0.01 | 0.04 |
| 5 | 1.67 | 5.94 |
| 6 | 2.67 | 4.27 |

### B. Probando la Dependencia de Largo Alcance (LRD)

Con la finalidad de analizar estadísticamente si una clase de tráfico particular exhibe pruebas de comportamiento LRD, se llevan a cabo pruebas de varianza y análisis de autocorrelación.

Para cada clase la Tabla II muestra el porcentaje de muestras y el volumen de tráfico correspondiente para diferentes rangos de $H$. En tanto que la Tabla III muestra los flujos de tráficos en términos de la actividad de red según los horarios de captura de tráfico anteriormente especificados, pero simplificados tan solo a tres componentes sustanciales actividad de red alta, actividad de red media y actividad de red baja (ver Subsección B de la Sección III).

TABLA II
PORCENTAJES DE MUESTRAS Y VOLÚMENES DE TRÁFICO INDICADOS PARA DIFERENTES VALORES DEL EXPONENTE DE HURST ($H$)

| Clase | Porcentaje de Muestras (%) | | | | Volumen de Tráfico (%) | | | |
|---|---|---|---|---|---|---|---|---|
|  | $H < 0.45$ | $0.45 < H < 0.5$ | $0.5 < H < 0.7$ | $H \geq 0.7$ | $H < 0.45$ | $0.45 < H < 0.5$ | $0.5 < H < 0.7$ | $H \geq 0.7$ |
| 1 | 76.5 | 0.0 | 17.6 | 5.9 | 99.3 | 0.0 | 0.6 | 0.1 |
| 2 | 50.0 | 6.7 | 30.0 | 13.3 | 37.5 | 3.3 | 43.7 | 15.5 |
| 3 | 17.2 | 6.9 | 20.7 | 55.2 | 4.9 | 6.6 | 7.5 | 81.0 |
| 4 | 91.7 | 8.3 | 0.0 | 0.0 | 98.9 | 1.2 | 0.0 | 0.0 |
| 5 | 40.9 | 18.2 | 22.7 | 18.2 | 37.0 | 30.4 | 19.0 | 13.7 |
| 6 | 20.7 | 21.3 | 18.0 | 40.0 | 20.3 | 17.7 | 16.0 | 46.0 |

TABLA III
PORCENTAJES DE MUESTRAS Y VOLÚMENES DE TRÁFICO CON $H > 0.5$ PARA DIFERENTES PERÍODOS DE ACTIVIDAD DE LA RED

| Actividad de Red | Porcentaje de Muestras (%) | | Volumen de Tráfico (%) | |
|---|---|---|---|---|
|  | $0.5 < H < 0.7$ | $H \geq 0.7$ | $0.5 < H < 0.7$ | $H \geq 0.7$ |
| Alta | 14.3 | 78.6 | 4.5 | 95.5 |
| Media | 33.3 | 66.7 | 1.6 | 98.4 |
| Baja | 22.2 | 11.1 | 25.6 | 14.3 |

El análisis de las Tablas II y III exhibe que las distintas clases pueden comportarse de manera muy diferente. Considérese que la mayor parte del volumen de tráfico, la **Clase 3**, se incluye en estas muestras. La **Clase 3** (tráfico HTTP) es claramente la que muestra el mayor grado de impulsividad. La mayoría de las muestras (76%) exhiben $H > 0.5$ y el 55% de ellas $H > 0.7$.

Es notorio que $H$ aumenta con la actividad de la red, lo cual es consistente con [7]. Aunque las Tablas II y III anteriores no diferencian los resultados por interfaz, el análisis de tráfico por interfaz muestra la misma tendencia.

Excluyendo la **Clase 3**, la relación entre $H$ y los volúmenes de tráfico no es clara para las clases restantes que pueden indicar una dependencia del tipo de aplicación. De hecho la **Clase 1** se comporta de manera opuesta y la **Clase 4** no muestra evidencia de impulsividad para los diferentes períodos de actividad. Esto puede deberse a la naturaleza regular del tráfico que atraviesa, como por ejemplo el tráfico de enrutamiento. Con respecto a la autocorrelación casi todas las muestras presentan funciones de autocorrelación que descienden lentamente a cero (lo cual es un indicativo de LRD) y tienen un $H$ mayor que 0.5.

## V. Conclusiones

En este estudio el parámetro Hurst se utiliza para medir LRD en muestras reales de tráfico clasificadas de acuerdo con un criterio propuesto. Los valores para $H$ se determinan para las clases de tráfico definidas y para los períodos de actividad.

Los resultados muestran que las **Clases 2** y **3**; transferencia masiva y tráfico HTTP, respectivamente juegan un papel muy importante en la carga total por interfaz. En particular la **Clase 3** es claramente la que muestra mayor evidencia de impulsividad, la cual aumenta con la carga de tráfico.

Mientras que la **Clase 2** tiene características autosimilares, la **Clase 1** se comporta de manera opuesta. La **Clase 4** presenta un $H$ estimado por debajo de 0.5 independientemente del período de actividad de la red. Para la mayoría de las muestras, las pruebas ilustran resultados similares ya sea analizando las series temporales de paquetes o bytes.